\documentclass[prb,aps,twocolumn,amsmath,amssymb,showpacs]{revtex4}
\usepackage{graphicx}
\usepackage{psfrag}
\usepackage{color}
\usepackage{subfigure}
\usepackage{amsmath}
\definecolor{dred}{rgb}{0.7,0.0,0.0}
\allowdisplaybreaks 
\usepackage{array}

\newcommand{\BE}{\begin{equation}}
\newcommand{\BEA}{\begin{eqnarray}}
\newcommand{\EE}{\end{equation}}
\newcommand{\EEA}{\end{eqnarray}}

\newcommand{\e}{{\rm e}}

\begin{document}

%
%

\title{
Magnetic properties and Mott transition in the square-lattice Hubbard model with frustration
}
\author{A. Yamada$^1$, K. Seki$^1$, R. Eder$^{2}$, and Y. Ohta$^1$}
\affiliation{$^1$Department of Physics, Chiba University, Chiba 263-8522, Japan\\
$^2$Karlsruhe Institut of Technology,
Institut f\"ur Festk\"orperphysik, 76021 Karlsruhe, Germany}

\date{\today}

\begin{abstract}

The magnetic properties and Mott transition of the Hubbard model 
on the square lattice with frustration are studied at half-filling and zero temperature by 
the variational cluster approximation. 
When the on-site repulsion $U$ is large, magnetically disordered state is realized in highly frustrated region 
between the N\'eel and collinear phases, and no imcommensurate magnetic states are found there. 
As for the Mott transition, in addition to the Mott gap and double occupancy, which clarify the nature of the transition, 
the structure of the self-energy in the spectral representation is studied in detail 
below and above the Mott transition point. 
The spectral structure of the self-energy is almost featureless in the metallic phase, 
but clear single dispersion, leading to the Mott gap, appears in the Mott insulator phase. 

\end{abstract}
 
 
\maketitle

%
%
\section{Introduction}
\begin{figure}
\includegraphics[width=0.47\textwidth,trim = 0 0 0 0,clip]{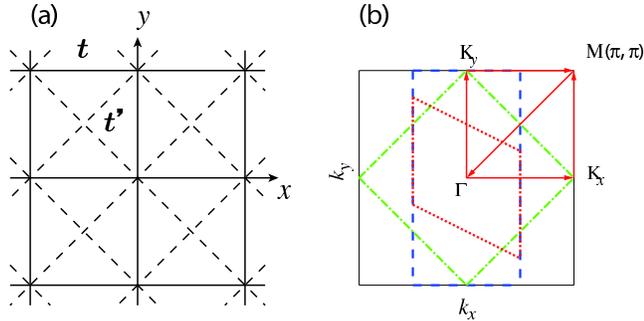}
\caption{
(a) Lattice structure of the Hubbard model 
on the square lattice with nearest- and next-nearest-neighbor hoppings $t$ and $t'$.
(b) The first Brillouin zone of the square lattice (solid outer square). 
The magnetic Brillouin zones corresponding to the 
AF, AF2, and AFC orderings in Fig.~\ref{fig:spin} are also depicted 
with dash-dotted (AF), dotted (AF2), and dashed (AFC)
lines, respectively. 
\label{fig:model}
}
\end{figure}
Geometric frustration is one of the most important issues in strongly correlated electron systems.
In addition to the experimental discovery of the high $T_c$ cuprates,\cite{bednors86} for example, 
the triangular-lattice organic materials $\kappa$-(BEDT-TTF)$_2\mathrm{X}$ \cite{lefebvre00,shimizu03,kagawa04} 
and herbertsmithite $\mathrm{ZnCu3(OH)_6Cl_2}$ \cite{helton07,mendels07} with kagom\'e 
lattice structure have inspired a lot of theoretical interest. In fact  
as a theoretical model incorporating the strong electron correlations and geometric frustrations, 
the Hubbard model on the square, triangle, and kagom\'e 
lattices is actively studied using non-perturbative methods, 
\cite{Senechal:2005,Aichhorn:2005,yokoyama,Aichhorn:2007,mizusaki,nev,yoshikawa,yamaki,sahebsara06,kyung06,ohashi08,
liebsch09,imai03,bulut05,ohashi06,udagawa10,kawakami10,atsushi11}
aiming to shed lights on the role of these effects on magnetism, Mott physics, and superconductivity. 

\begin{figure}
\includegraphics[width=0.47\textwidth,trim = 0 0 0 0,clip]{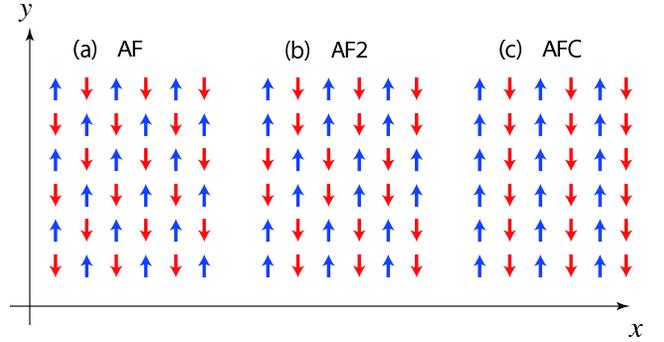}
\caption{
The spin configurations of the (a) AF (N\'eel) , 
(b) AF2, and (c) AFC (collinear) states. 
\label{fig:spin}\\[-1.5em]
}
\end{figure}
In this paper, we study the magnetic properties and Mott transition in the Hubbard model 
on the square lattice with nearest- and next-nearest-neighbor hoppings $t$ and $t'$, and on-site Coulomb repulsion $U$ 
(see Fig.~\ref{fig:model}) 
at half-filling and zero temperature using the variational cluster approximation 
(VCA),\cite{Senechal00,Potthoff:2003-1,Potthoff:2003} 
which is formulated based on a rigorous variational principle and exactly takes into account the short-range correlations. 

\begin{figure*}
\includegraphics[width=0.77\textwidth,trim = 0 0 0 0,clip]{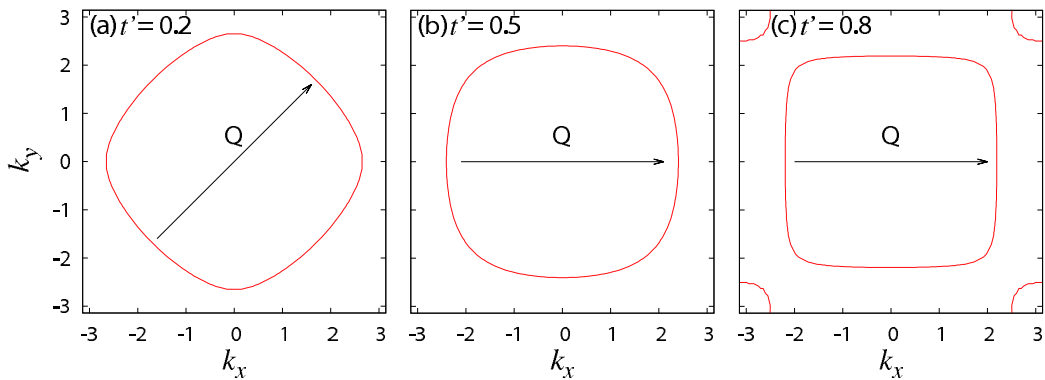}
\caption{
(Color online)
The noninteracting ($U=0$) Fermi surface at half-filling for (a) $t' = 0.2$, (b) $t' = 0.5$, and (c) $t' = 0.8$. 
The best nesting vector $Q$ is also shown. 
The change in the topology (Lifshitz transition) takes place around $t' \simeq 0.71$. 
\label{fig:fs}
}
\end{figure*}
The ordinary N\'eel state with the ordering wave vector $Q=(\pi,\pi)$, 
the so-called collinear phase with $Q=(\pi,0)$, 
and a third state with longer periodicity in real space corresponding to $Q=(\pi,\pi/2)$ 
are considered for the magnetic orderings 
and 12-site clusters of the three different shapes 
are used to see the finite-size effects of the calculations. 
Hereafter, the magnetic orderings with $Q=(\pi,\pi)$,  $(\pi,\pi/2)$, and  $(\pi,0)$ 
are referred to as AF, AF2, and AFC, and they are illustrated in Fig.~\ref{fig:spin}. 

We find that, when $U$ is larger than the band width $W=8t$, 
the paramagnetic (PM) phase is realized between the AF ($t'/t \lesssim 0.65 $) and AFC ($0.8 \lesssim t'/t$) phases, 
and there is no area where the AF2 phase is realized. 
For $ U \gtrsim  W $, the cluster-shape dependence of the calculations is small and 
our results agree well with the analysis of the Heisenberg model.
\cite{Schulz92-96,j1j2-2,j1j2-3,dimer1,dimer2,dimer3,j1j2-4,j1j2-5,j1j2-6,j1j2-7,j1j2-8} 
in the large $U$ limit. 
For $U \lesssim W$, VCA on the 3$\times$4 cluster shows that there is a small region in the phase diagram 
where the AF2 phase is realized ($0.7 \lesssim t'/t \lesssim 0.85$ and $4.5  \lesssim U/t \lesssim 6.5$), which qualitatively 
agrees with the mean-field approximation, though the finite-size effects turn out to be non-negligible. 

As for the Mott transition, in addition to the Mott gap and double occupancy, 
which clarify the nature of the transition, 
we study in detail the spectral structure of the self-energy in its spectral representation.\cite{l1,eder1} 
We found that the Mott transition takes place below the band width and is of the second order. 
The spectral structure of the self-energy is almost featureless in the metallic phase, 
but clear single dispersion, leading to the Mott gap, appears in the Mott insulator phase.

%
%
\section{The magnetic properties in the mean-field approximation}
\label{sec:model}

The Hamiltonian of the 
Hubbard model on the square lattice 
with nearest- and next-nearest-neighbor hoppings $t$ and $t'$, and the on-site Coulomb repulsion $U$ 
reads 
\begin{align}
H = -\sum_{i,j,\sigma} t_{ij}c_{i\sigma }^\dag c_{j\sigma}
+ U \sum_{i} n_{i\uparrow} n_{i\downarrow} - \mu \sum_{i,\sigma} n_{i\sigma},
\label{eqn:hm}
\end{align}
where $t_{ij}=t$ if the sites $i$ and $j$ are nearest-neighbor, $t_{ij}=t'$ if they are next-nearest-neighbor, 
and $t_{ij}=0$ otherwise, and $\mu$ is the chemical potential. 
The annihilation (creation) operator for an electron at site $i$ with spin $\sigma$ is denoted as 
$c_{j\sigma}$ ($c_{i\sigma }^\dag$) and $n_{i\sigma}=c_{i\sigma}^\dag c_{i\sigma}$. 
The hopping $t'$ introduces frustration to the N\'eel ordering since it yields the antiferromagnetic interactions 
between next-nearest-neighbor spins, which are parallel in the N\'eel ordering. The energy unit is set as $t=1$ hereafter. 

The energy band for the noninteracting case ($U=0$) is 
\begin{align}
\varepsilon_k = -2 t \sum_{i=x,y} \cos k_i  -4t'\cos k_x \cos k_y  - \mu,
\label{eqn:band-free}
\end{align}
$(- \pi < k_i < \pi)$, so the band width is $W = 8t$. 
The noninteracting Fermi surface at half-filling is depicted in Fig.~\ref{fig:fs} 
for (a) $t' = 0.2$, (b) $t' = 0.5$, and (c) $t' = 0.8$ 
together with the best nesting vector $Q$.

First, we briefly study the magnetic properties of this model at half-filling in the large $U$ expansion. 
The leading order approximation of this expansion 
leads to the effective spin Hamiltonian ($J_1-J_2$ Heisenberg model)
\begin{align}
H = J_1 \sum_{i,j} S_i \cdot S_j   +  J_2 \sum_{i,j} S_i \cdot S_j,
\label{eqn:hj1j2}
\end{align}
where $J_1 = 4t^2/U$ and $J_2 = 4t'^2/U$, 
and the next leading corrections to the Hamiltonian (\ref{eqn:hj1j2}) are of the order $1/U^3$, 
as are calculated e.g. in Refs.~\onlinecite{expansion-1,expansion-2,expansion-3}. 
The (classical) energy per site of the Hamiltonian (\ref{eqn:hj1j2}) for the magnetic orderings with $Q=(\pi,\pi/n)$ 
($n=1,2, \cdots $) is evaluated as 
\begin{equation}
E_n = - \frac{1}{2}\{ J_2 + \frac{1}{n}(J_1 - 2  J_2 ) \}, \nonumber 
\end{equation}%
where $n = 1$ for AF, $n =2$ for AF2, and $n \rightarrow \infty$ for AFC. 
Therefore at $t' = t'_c = t/\sqrt{2} \simeq 0.707t $ all the $E_{n}$ are exactly degenerate in this approximation, and 
$E_{\rm AF} < E_{\rm AF2} < \cdots < E_{\rm AFC}$ for $t' < t'_c$ and 
$E_{\rm AF} > E_{\rm AF2} > \cdots > E_{\rm AFC}$ for $t' > t'_c$.

\begin{figure}
\includegraphics[width=0.45\textwidth,trim = 0 0 0 0,clip]{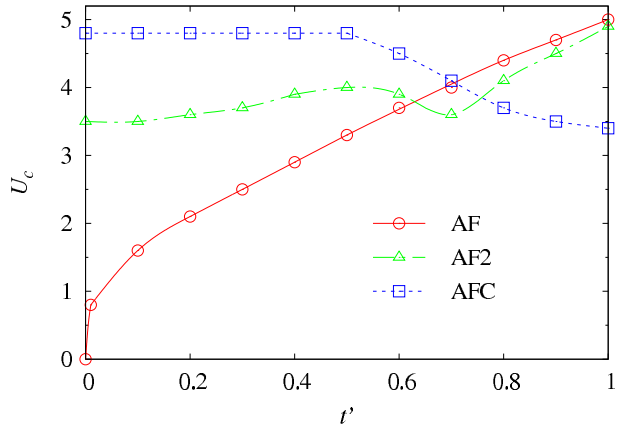}\\[-0.5em]
\caption
{
(Color online) 
The critical  interaction strength $U_c$ as a function of $t'$ 
for the AF (circles), AF2 (triangles), and AFC (squares) orderings in the mean-field approximation. 
Nontrivial solutions are obtained for $U \geq U_c$. 
Lines are guides to the eye. 
\label{fig:uc}\\[-2.5em]
}
\end{figure}
Next we study the magnetic properties of this model in the mean-field approximation 
taking into account the three magnetic states AF, AF2, and AFC. 
The Hamiltonian in the mean-field approximation is constructed following the usual procedure 
and we set the average number $\langle n_{i\sigma} \rangle$ as 
\begin{equation}
\langle n_{i\sigma} \rangle = \frac{1}{2}n + \frac{1}{2}\sigma {\rm sign}(i)\Delta,
\nonumber 
\end{equation}
where ${\rm sign}(i) = +1\,\,\, (-1)$ for the spin up (down) sites in Fig.~\ref{fig:spin}, $n =1$ (at half-filling), 
and the parameter $\Delta$ becomes the order parameter per site for the consistent mean-field solutions. 
The critical interaction strength $U_c$, above which nontrivial solutions are obtained,  
is calculated at zero temperature as a function of $t'$ in Fig.~\ref{fig:uc} 
for the AF (circles), AF2 (triangles), and AFC (squares) orderings. 
\begin{figure}
\includegraphics[width=0.3\textwidth,trim = 0 0 0 0,clip]{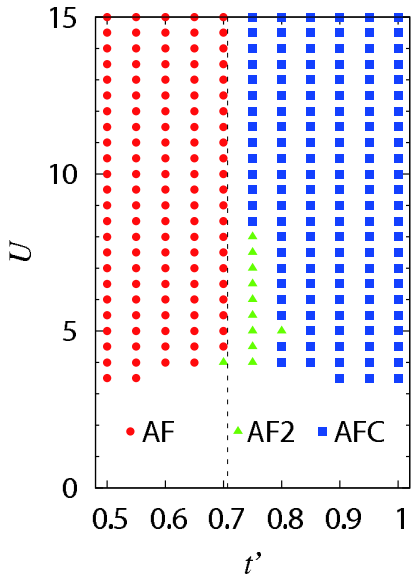}
\caption
{
(Color online) 
Phase diagram at half-filling and zero temperature as a function of $t'$ and $U$ in the mean 
field approximation. 
The circles, triangles, and squares represent the AF, AF2, and AFC states, respectively. 
In the classical approximation of the $J_1-J_2$ Heisenberg model (see text), 
the energies of the AF, AF2, and AFC states are exactly degenerate on the vertical dotted line 
($t' = t'_{c} \simeq 0.71$)
and the AF state is stable for $t' < t'_{c}$ while the AFC state is realized for $t'_{c} < t'$. 
\label{fig:phase-diagram-mean}
}
\end{figure}

The phase diagram is calculated as a function of $t'$ and $U$ in Fig.~\ref{fig:phase-diagram-mean} 
by comparing the energies of the magnetic and PM states. 
At $t' = 0.5$ the best nesting vector seems to be $Q=(\pi,0)$ but according to the 
mean-field analysis the AF state is realized. 
In the large $U$ region 
the mean-field analysis agrees with the classical approximation of the $J_1-J_2$ Heisenberg model, 
which predicts that the AF and AFC phases are separated by the vertical dotted line ($t' = t'_{c}$) 
in Fig.~\ref{fig:phase-diagram-mean}. 
Our results are consistent with the previous mean-field analyses\cite{hofstetter,yu-yin} 
studying the AF and AFC states. 

%

\section{Variational cluster approximation}
\label{sec:VCA}

\begin{figure*}
\includegraphics[width=0.8\textwidth,trim = 0 0 0 0,clip]{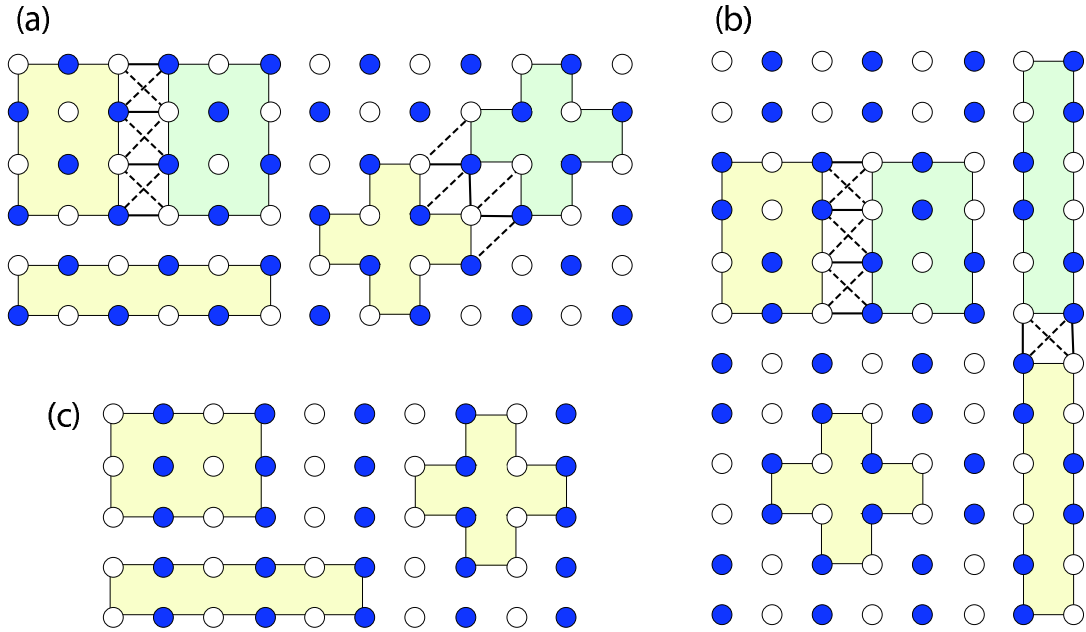}
\caption
{
(Color online)
The clusters used to investigate the (a) AF, (b) AF2, and (c) AFC states.
When necessary, these 12-site clusters are paired with their rotated or reflected mirror images 
so that the extended 24-site clusters tile the lattice in the Bravais sense in the presence of the Weiss fields. 
\label{fig:cluster}
}
\end{figure*}
Next we apply VCA\cite{Senechal00,Potthoff:2003-1,Potthoff:2003} to 
examine the magnetic properties and Mott transitions of this model. 
VCA is an extension of the cluster perturbation theory\cite{Senechal00} based on the 
self-energy-functional approach.\cite{Potthoff:2003} 
This approach uses the rigorous variational principle 
\begin{equation}
\delta \Omega _{\mathbf{t}}[\Sigma ]/\delta \Sigma =0
\label{eq:dyson-eq}
\end{equation}%
for the thermodynamic grand-potential 
$\Omega _{\mathbf{t}}$ written 
in the form of a functional of the self-energy $\Sigma $ as 
\begin{equation}
\Omega _{\mathbf{t}}[\Sigma ]=F[\Sigma ]+\mathrm{Tr}\ln
(-(G_0^{-1}-\Sigma )^{-1}),
\label{eq:f-sigma}
\end{equation}%
where $F[\Sigma ]$ is the Legendre 
transform of the Luttinger-Ward functional\cite{lw} and the index $\mathbf{t}$ denotes the explicit dependence of 
$\Omega _{\mathbf{t}}$ on all the one-body operators in the Hamiltonian. 
The stationary condition (\ref{eq:dyson-eq}) for $\Omega_{\mathbf{t}}[\Sigma ]$ leads to the Dyson's equation. 
All Hamiltonians with the same interaction part share the same functional form of $F[\Sigma ]$, and using that property 
$F[\Sigma ]$ can be evaluated from the exact solution of a simpler Hamiltonian $H'$, though 
the space of the self-energies where $F[\Sigma ]$ is evaluated is now restricted to that of $H'$. 
In VCA, one uses for $H'$ a Hamiltonian formed of clusters that are disconnected by removing hopping terms 
between identical clusters that tile the infinite lattice. 
A possible symmetry breaking is investigated by including 
into $H'$ the corresponding Weiss field that will be determined by minimizing the 
grand-potential $\Omega _{\mathbf{t}}$. 
Rewriting $F[\Sigma ]$ in Eq. (\ref{eq:f-sigma}) 
in terms of the grand-potential $\Omega'\equiv \Omega'_\mathbf{t}[\Sigma]$ and 
Green function $G'{}^{-1}\equiv G'_0{}^{-1}-\Sigma$ of the cluster Hamiltonian $H'$, the grand-potential becomes a 
function of $\mathbf{t}'$ expressed as 
\begin{equation}
\Omega _{\mathbf{t}}(\mathbf{t}')=\Omega'\kern-0.4em - \kern-0.4em\int_C{\frac{%
d\omega }{2\pi }} \e^{ \delta \omega} \sum_{\mathbf{K}}\ln \det \left(
1+(G_0^{-1}\kern-0.2em -G_0'{}^{-1})G'\right)
\label{omega}, 
\end{equation}%
where the functional trace has become an integral over the diagonal variables 
(frequency and super-lattice wave vectors) of the logarithm of a determinant over intra-cluster indices. 
The frequency integral 
is carried along the imaginary axis and $\delta \rightarrow + 0$. 
The stationary solution of $\Omega _{\mathbf{t}}(\mathbf{t}')$ and the 
exact self-energy of $H'$ at the stationary point, denoted as $\Sigma^{*}$, are the approximate grand-potential 
and self-energy of $H$ 
in VCA, and physical quantities, such as expectation values of the one-body operators, are calculated using 
the Green function $G_0{}^{-1}-\Sigma^{*} $. 
In VCA, the restriction of the space of the self-energies $\Sigma$ into that of $H'$ 
is the only approximation 
involved and the electron correlations within the cluster used to set up  $H'$ 
are rigorously taken into account by exactly solving $H'$.

\begin{figure*}
\includegraphics[width=0.92\textwidth,trim = 0 0 0 0,clip]{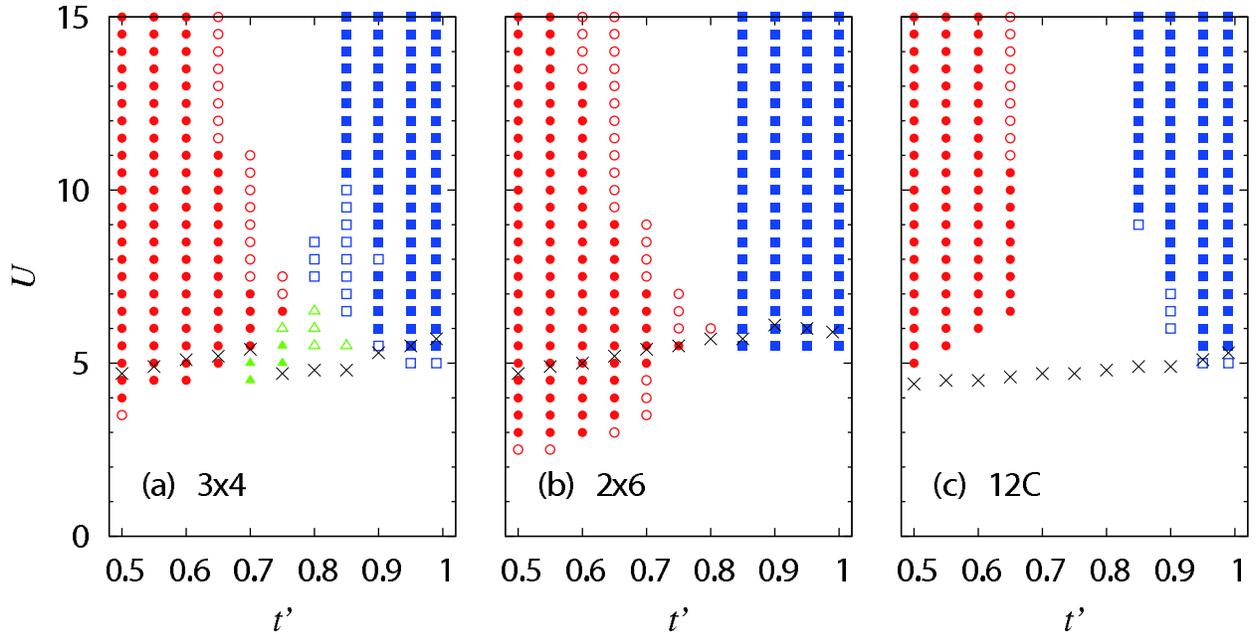}\\[-0.5em]
\caption{
(Color online)
The phase diagram at half-filling and zero temperature 
as a function of $t'$ and $U$ obtained by VCA using the (a) 3$\times$4, (b) 2$\times$6, and (c) 12C cluster. 
The circles, triangles, and squares represent the AF, AF2, and AFC states, respectively. 
Non-filled marks indicate that their energies are degenerate with other solutions within the accuracy of $10^{-3}$. 
The crosses are the Mott transition points obtained in each cluster assuming that no magnetic order is allowed. 
\label{fig:phase-vca}\\[-0.5em]
}
\end{figure*}
In our analysis, the three magnetic orderings AF, AF2, and AFC are considered as possible symmetry 
breaking patterns, and the 12-site clusters of the three different shapes depicted in Fig.~\ref{fig:cluster} 
(3$\times$4, 2$\times$6, and a cross-shaped 12-site cluster, referred to as 12C hereafter)
are used to set up the cluster Hamiltonian $H'$, 
where the Weiss field 
\begin{eqnarray}
H_{\rm M}&=& h_{\rm M}\sum_{i} {\rm sign}(i)(n_{i\uparrow }-n_{i\downarrow }) 
\label{weiss}
\end{eqnarray}
with ${\rm sign}(i) = +1,\,\,\, (-1)$ for the shaded (white) sites in Fig.~\ref{fig:cluster} is included. 

As is shown in Fig.~\ref{fig:cluster}, when necessary (3$\times$4 and 12C in (a), and 3$\times$4 and 2$\times$6 in (b)), 
these 12-site clusters are paired 
with their appropriate rotated or reflected images so that the extended 24-site clusters 
tile the infinite lattice in the Bravais sense in the presence of the magnetic ordering. 
In these cases the Green function $G'$ in Eq. (\ref{omega}) is given by
\begin{eqnarray}
G'^{-1}&=&{G'}_1^{-1} + {G'}_2^{-1} + t_{12}
\end{eqnarray}
where ${G'}_1$ is the exact Green function on a 12-site cluster (the site and spin indices suppressed), 
${G'}_2$ is the exact Green function of the mirror image (a simple transformation of ${G'}_1$), 
and $t_{12}$ is the hopping matrix linking the two 12-site clusters (the solid and dashed links in Fig.~\ref{fig:cluster}). 
In all cases, the correlations within the 3$\times$4, 2$\times$6, and 12C clusters are rigorously taken into account 
by exactly diagonalizing $H'$. 
The cluster-shape dependence of the results is a measure of the finite-size effects in our analysis. 
\begin{figure*}
\includegraphics[width=0.92\textwidth,trim = 0 0 0 0,clip]{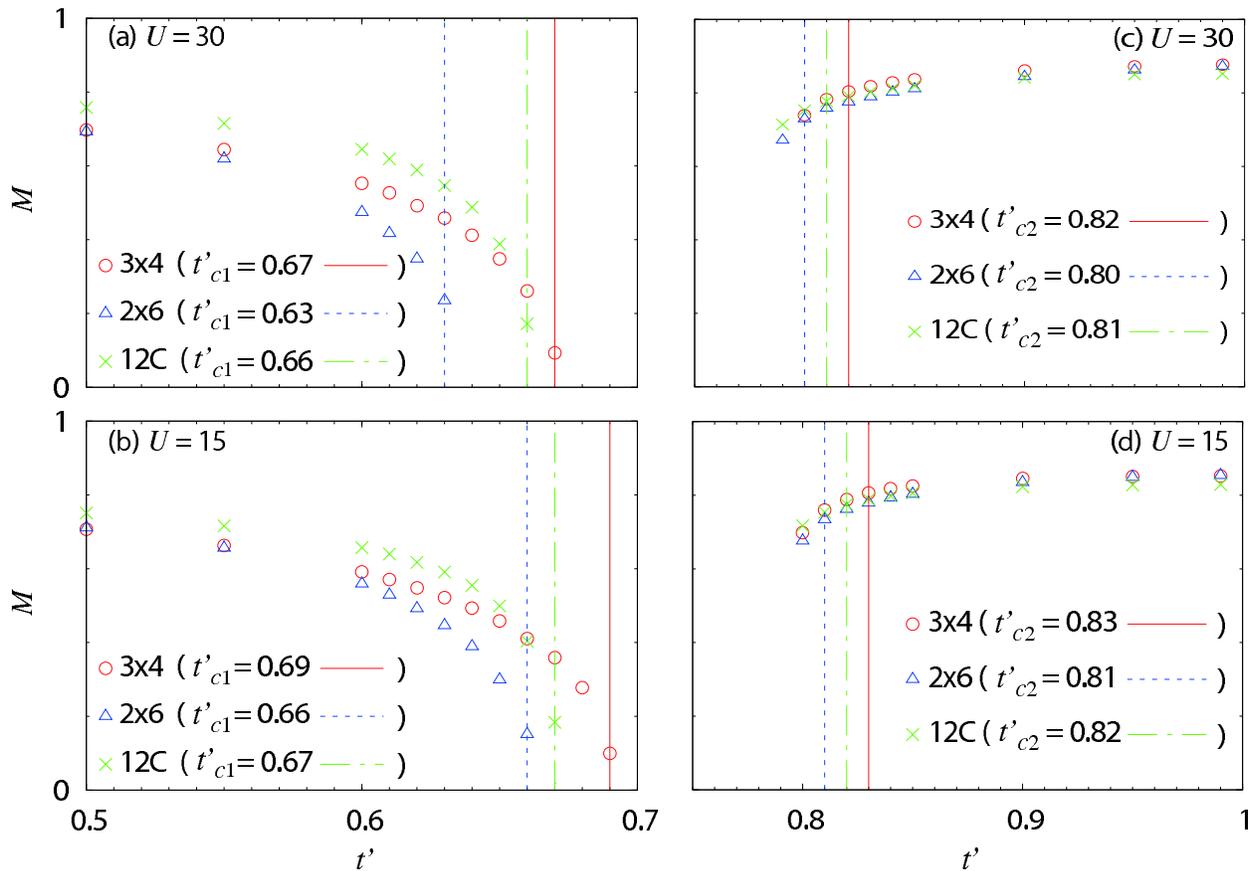}
\caption{
(Color online)
The order parameter of the AF state at (a) $U = 30$ and (b) $U = 15$ and 
that of the AFC state at (c) $U = 30$ and (d) $U = 15$ as a function of $t'$ 
obtained by VCA using the 3$\times$4 (circles), 2$\times$6 (triangles), and 12C (crosses) clusters.
The critical couplings $t'_{c1}$ separating the AF and PM phases 
and $t'_{c2}$ separating the AFC and PM phases are also shown 
by the vertical full (3$\times$4), dotted (2$\times$6), and dash-dotted (12C) lines. 
In (c) and (d) the data points for $t' < t'_{c2}$ are energetically disfavored since their energies are higher 
than the PM solutions by more than $10^{-3}$.
\label{fig:af-afs-order-u}
}
\end{figure*}
With this set up, the grand-potential per site $\Omega$ is now a function 
$\Omega(\mu', h_{\rm M})$ of the Weiss field $h_{\rm M}$ 
and the cluster chemical potential $\mu'$, where the latter should be included 
for the thermodynamic consistency,\cite{aichhorn} 
and the variational principle (\ref{eq:dyson-eq}) becomes the stationary point search of 
$\Omega(\mu', h_{\rm M})$ with respect to $h_{\rm M}$ and $\mu'$. 
During the search, the chemical potential of the system $\mu$ is also adjusted so that the electron 
density per site $n$ is equal to 1 within the accuracy of $10^{-4}$ (i.e. $|n-1| < 10^{-4}$). 
In general, a stationary solution with $h_{\rm M} \neq 0$ corresponding to the magnetically ordered state and 
that with $h_{\rm M} = 0$ corresponding to the PM state are obtained, and the energies per site 
$E=\Omega+\mu n$ are compared for the AF, AF2, AFC, and PM states to determine the ground state. 
The density of state per site 
\begin{eqnarray}
D(\omega)= \lim_{\eta \rightarrow 0}  \int
{\frac{%
d^2 k }{(2\pi)^2 }}\frac{1}{n_c}\sum_{\sigma, a=1}^{n_c}\{ -\frac{1}{\pi} \mathrm{Im}G_{a\sigma}(k, \omega+i\eta) \}
\label{eq:dos}
\end{eqnarray}
is also calculated to examine the gap, where 
$n_c$ is the number of the sites in the unit cell in the sense of the sub-lattice formalism with the magnetic orderings 
($n_c = 4$ for the AF2 state, $n_c = 2$ for the AF and AFC states, and $n_c = 1$ for the PM state), 
and the $k$ integration is over the corresponding Brillouin zone (see Fig.~\ref{fig:model} (b)). 
In Eq. (\ref{eq:dos}), $\eta \rightarrow 0$ limit is evaluated using the standard extrapolation method 
by calculating $D(\omega)$ for $\eta =0.1$, $0.05$, and $0.025$. 
The numerical error after this extrapolation is estimated to be of order $10^{-3}$, 
so the gap is identified as the region of $\omega$ around $\omega \simeq 0 $ 
where the extrapolated $D(\omega)$ is less than $10^{-2}$. 
We also compute the magnetic order parameter per site 
\begin{eqnarray}
M&=& \frac{1}{n_c}\sum_{a=1}^{n_c} {\rm sign}(a)( \langle n_{a\uparrow } \rangle - \langle n_{a\downarrow } \rangle ) \nonumber
\end{eqnarray}
and the double occupancy per site 
\begin{eqnarray}
D_{\rm occ}= \frac{1}{n_c}\sum_{a=1}^{n_c} \langle n_{a\uparrow } n_{a\downarrow } \rangle = \frac{dE}{dU} \nonumber 
\end{eqnarray}
where $ \langle n_{a\sigma } \rangle $ and $\langle n_{a\uparrow } n_{a\downarrow } \rangle$ 
are the expectation values of $n_{a\sigma }$ and 
$n_{a\uparrow } n_{a\downarrow }$, respectively.

\section{Magnetism and Mott transition}
\label{sec:results}

\subsection{Phase diagram}

Fig.~\ref{fig:phase-vca} shows the phase diagram at half-filling and zero temperature obtained 
by VCA on the (a) 3$\times$4, (b) 2$\times$6, and (c) 12C cluster, where 
the circles, triangles, and squares represent the AF, AF2, and AFC states, respectively. 
The numerical error of the ground-state energy per site due to the error of the number density $n$ is at most 
of the order $10^{-4}$, 
so when the ground-state energy of a magnetic state is degenerate with that of other states (e.g. PM state) 
within the accuracy of $10^{-3}$ we have used non-filled circles, triangles, and squares to indicate it. 
(Later we call a solution as energetically disfavored 
if its energy is higher than the ground-state energy by more than $10^{-3}$.)
The crosses are the Mott transition points obtained in each cluster assuming 
that no magnetic order is allowed (i.e. $h_{\rm M}$ = 0). 
As is mentioned earlier, the cluster-shape dependence of the results is a measure of the finite-size effects in our analysis. 

In the region $ U \gtrsim  W (=8) $, the results of the three clusters agree rather well, and 
for $t' \lesssim 0.65$ the AF state is realized, while $0.8 \lesssim  t'$ the AFC state is stable, 
and the PM state is realized between the AF and AFC states. 
There is no region where the incommensurate AF2 phase is realized for $ U \gtrsim  W $, 
so our VCA result disagrees with that of Ref. \onlinecite{mizusaki}. 
When $ U \lesssim W $, the three clusters give qualitatively different results and  
the cluster-shape dependence is not negligible. 
In this area, the 3$\times$4 cluster 
yields the results qualitatively very similar to the 
mean-field approximation in Fig.~\ref{fig:phase-diagram-mean}, and shows 
that the AF2 state is realized in the region $0.7 \lesssim t' \lesssim 0.85$ and $4.5 \lesssim U \lesssim 6.5$. 
Other incommensurate states may be also realized around this area.

The Mott transition takes place for $4 \lesssim U \lesssim 6$.
Our critical interaction strength $U_{\rm MI}$ is larger than the results 
of VCA on the clusters up to 8 sites.\cite{nev} 
In general, as a cluster becomes larger the kinetic energies simulated on it increase as an average, 
and $U_{\rm MI}$ increases. 
Our result is consistent with this observation.

\subsection{Magnetic properties}

Now we study in detail the magnetic properties and first consider the strong coupling region. 
In general VCA gives the exact results regardless of the cluster size and shape 
when inter-site hopping interactions are set to be zero (i.e. $t=t'=0$ in the present model).
But it does not necessarily mean that VCA is a good approximation in the 
strong coupling limit ( $U \gg t,t' $) even for small clusters. 
In fact, in the present model, the phase structure is not determined by the ratio $t/U$ or $t'/U$, 
but is determined by the ratio $t/t'$ (or, $J_1 / J_2$ in the Heisenberg model) in the strong coupling region. 
Therefore the detailed analysis of this region is not a trivial check of the calculations. 
The order parameter of the AF and AFC states is computed as a function of $t'$ 
at $U=30$ and $U=15$ in Fig.~\ref{fig:af-afs-order-u} by VCA on the 3$\times$4 (circles), 2$\times$6 (triangles), and 
12C (crosses) clusters. The critical couplings $t'_{c1}$ separating the AF and PM phases
and $t'_{c2}$ separating the AFC and PM phases are also shown by 
the vertical full (3$\times$4), dotted (2$\times$6), and dash-dotted (12C) lines. 
In Fig.~\ref{fig:af-afs-order-u} (c) and (d) the data points for $t' < t'_{c2}$ correspond to energetically disfavored solutions 
because their energies are higher than the PM solutions by more than $10^{-3}$. 

As is seen in Fig.~\ref{fig:af-afs-order-u}, the cluster-shape dependence is rather small, especially at $U=30$. 
The transition between AF and PM states is of the second order since the order 
parameter smoothly goes to zero and there are no energetically disfavored AF solutions outside the 
AF phase. 
The transition between the AFC and PM states is first order (level crossing), 
since energetically disfavored AFC solutions exist outside the AFC phase. 
These nature of the phase transitions agree with the analysis of the Heisenberg model
\cite{j1j2-4,j1j2-5,j1j2-6,j1j2-7,j1j2-8} and VCA on 2$\times$2 cluster.\cite{yoshikawa,yamaki} 

The window of the PM phase between the AF and AFC phases is 
$t'_{c1} =0.63 \sim 0.67 < t' < t'_{c2} = 0.80 \sim 0.82$ at $U=30$ in our analysis. 
In the analysis of the Heisenberg model, the PM region appears in the range 
$0.4 \lesssim J_2 /J_1 \lesssim 0.6$, which translates into the region $0.63 \lesssim t' \lesssim 0.78 $ in the Hubbard model 
in the strong coupling limit and agrees reasonably well with our results. 
Among the three clusters, the window of the 2$\times$6 cluster is closest to the result of the 
Heisenberg model. 
This may be related to the proposal of Ref. \onlinecite{ueda} in the Heisenberg model 
that an array of spin singlets spontaneously formed on 2$\times$2 plaquettes is realized in the PM 
(magnetically disordered) ground state 
and only the 2$\times$6 cluster exactly contains three units of such 2$\times$2 plaquettes. 

Our window of the PM state is much wider than the VCA on the 8-site cluster,\cite{nev} 
mainly because our $t'_{c1}$ is much smaller than there. 
Since it is reported that spin structures extended to a few sites such as the dimer states\cite{dimer1,dimer2,dimer3} 
and spin singlets on 
2$\times$2 plaquettes\cite{ueda} will be realized in this PM region, it seems to be natural to expect that the PM 
region becomes wider for larger clusters, as is observed here. 
In addition to this discrepancy in the large $U$ limit, 
in Ref. \onlinecite{nev} the width of the PM state decreases as $U$ increases in the region $ U \gtrsim  W $, while 
in our result the width increases toward the large $U$ limit as $U$ increases. 
\begin{figure}
\includegraphics[width=0.47\textwidth,trim = 0 0 0 0,clip]{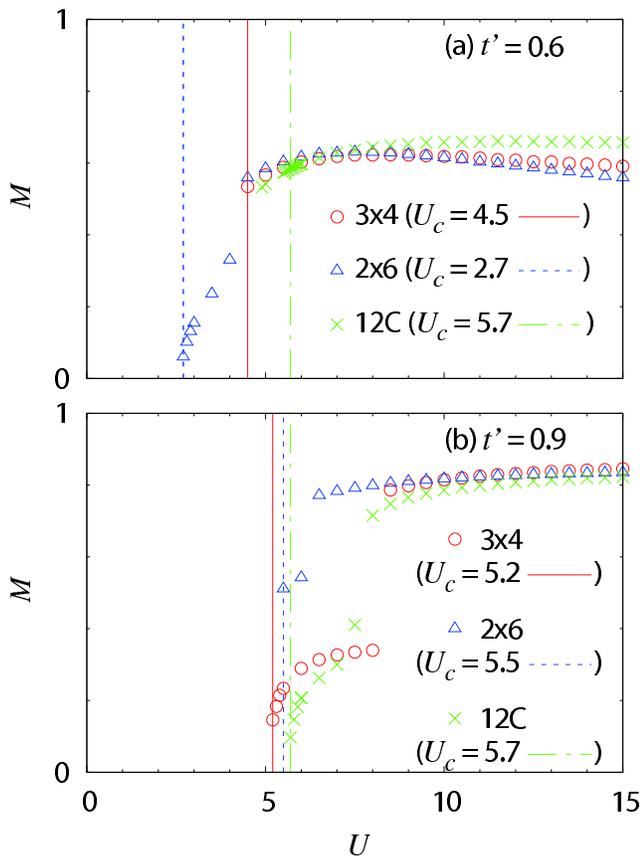}
\caption{
(Color online)
The order parameter of the (a) AF state at $t' = 0.6$ and (b) AFC state at $t' = 0.9$ as a function of $U$ 
obtained by VCA using the 3$\times$4 (circles), 2$\times$6 (triangles), and 12C (crosses) clusters.
The critical interaction strength $U_c$ separating the magnetic and PM phases 
phases is also shown by the vertical full (3$\times$4), dotted (2$\times$6), and dash-dotted (12C) lines. 
\label{fig:af-afs-order-06-09}
}
\end{figure}

The fact that our $t'_{c1}$ and $t'_{c2}$ in the large $U$ limit are close to the results of the Heisenberg model indicates 
that the correlations within and between composite structures of spins such as dimers 
and spin singlets on 2$\times$2 plaquettes 
are simulated well on the 12-site clusters for $ U \gtrsim  W $, which in turn 
supports the validity of our results about the absence of the AF2 state in that region 
since the AF2 ordering is also like a composite structure in the sense that its unit cell is four sites, 
and the analysis of the AF2 ordering requires simulating the correlations within and between these unit cells. 

Next we consider the region $U \lesssim W$. 
Fig.~\ref{fig:af-afs-order-06-09} shows 
the order parameter of the (a) AF state at $t' = 0.6$ and (b) AFC state at $t' = 0.9$ as a function of $U$ 
obtained by VCA using the 3$\times$4 (circles), 2$\times$6 (triangles), and 12C (crosses) clusters.
The critical interaction strength $U_c$ separating the magnetic and PM phases 
phases is also shown by the vertical full (3$\times$4), dotted (2$\times$6), and dash-dotted (12C) lines. 

The behavior of the AF order parameter is similar down to $U \simeq 5 $ 
for the three clusters, however in the case of the 12C cluster the AF solutions below $U < 5.7$ are energetically disfavored 
compared to the PM solutions so a first order transition takes place at $U_c = 5.7$. 
The other two clusters predict that the transition is of the second order since there are no energetically disfavored AF solutions 
below $U_c$. 

In the case of the AFC solutions, we found that there are typically two solutions, 
one with large $M$ and the other with smaller $M$. 
The energies of the two solutions cross around $U \simeq 6 \sim 8$, which results in a 
discontinuous change of the order parameter in Fig.~\ref{fig:af-afs-order-06-09} (b). 
The discontinuity of the AFC order parameter takes place (though at much smaller $U$) also 
in the mean-field approximation.\cite{yu-yin} 
All the three clusters predict that at $t'=0.9$ the transition between the AFC and PM states is of the second order 
since there are no energetically disfavored AFC solutions below $U_c$. 
\begin{figure}
\subfigure{\includegraphics[width=0.47\textwidth, trim = 70 80 250 680, clip]
 {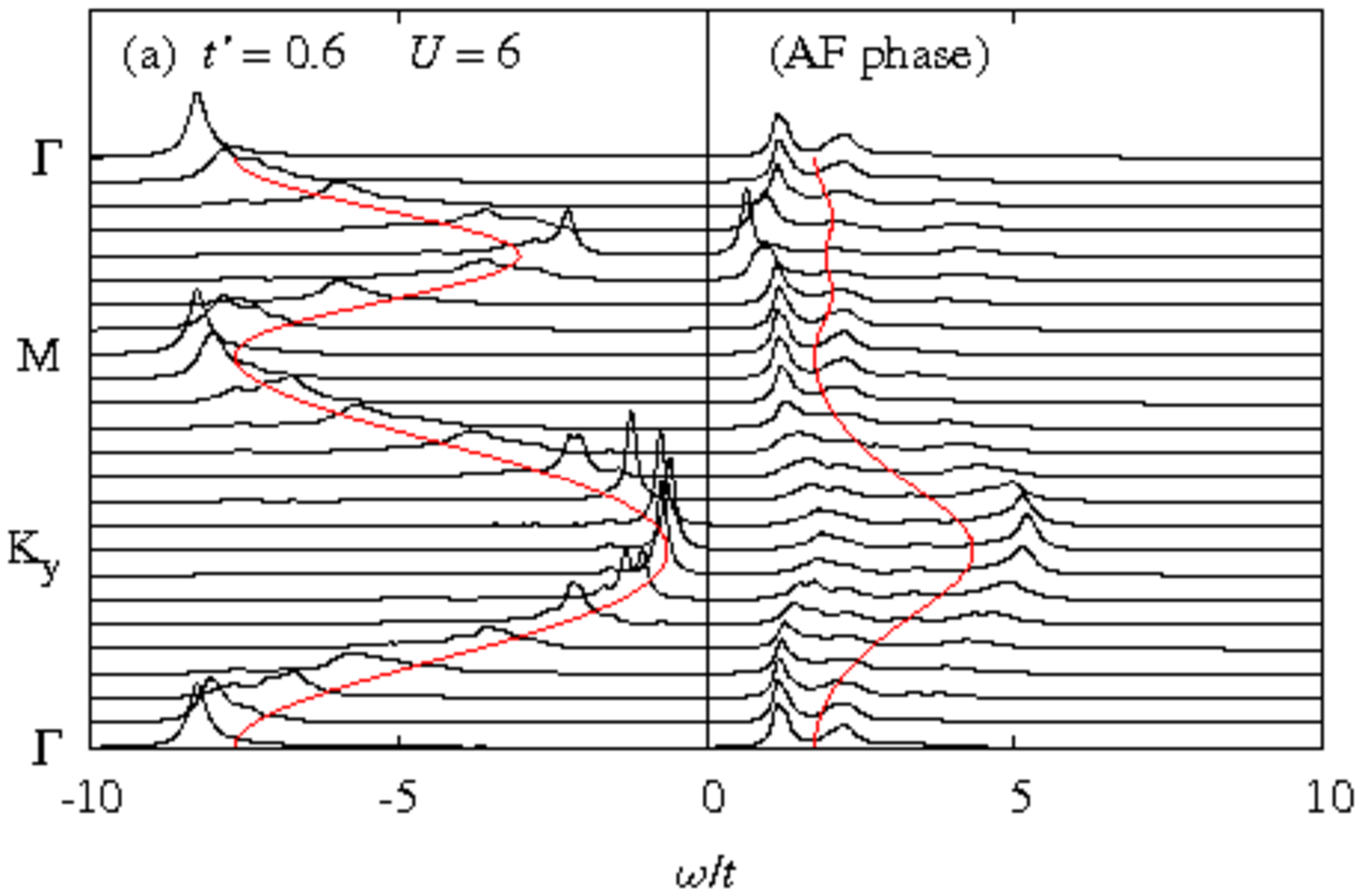}}
\subfigure{\includegraphics[width=0.47\textwidth, trim = 90 50 250 660, clip]
 {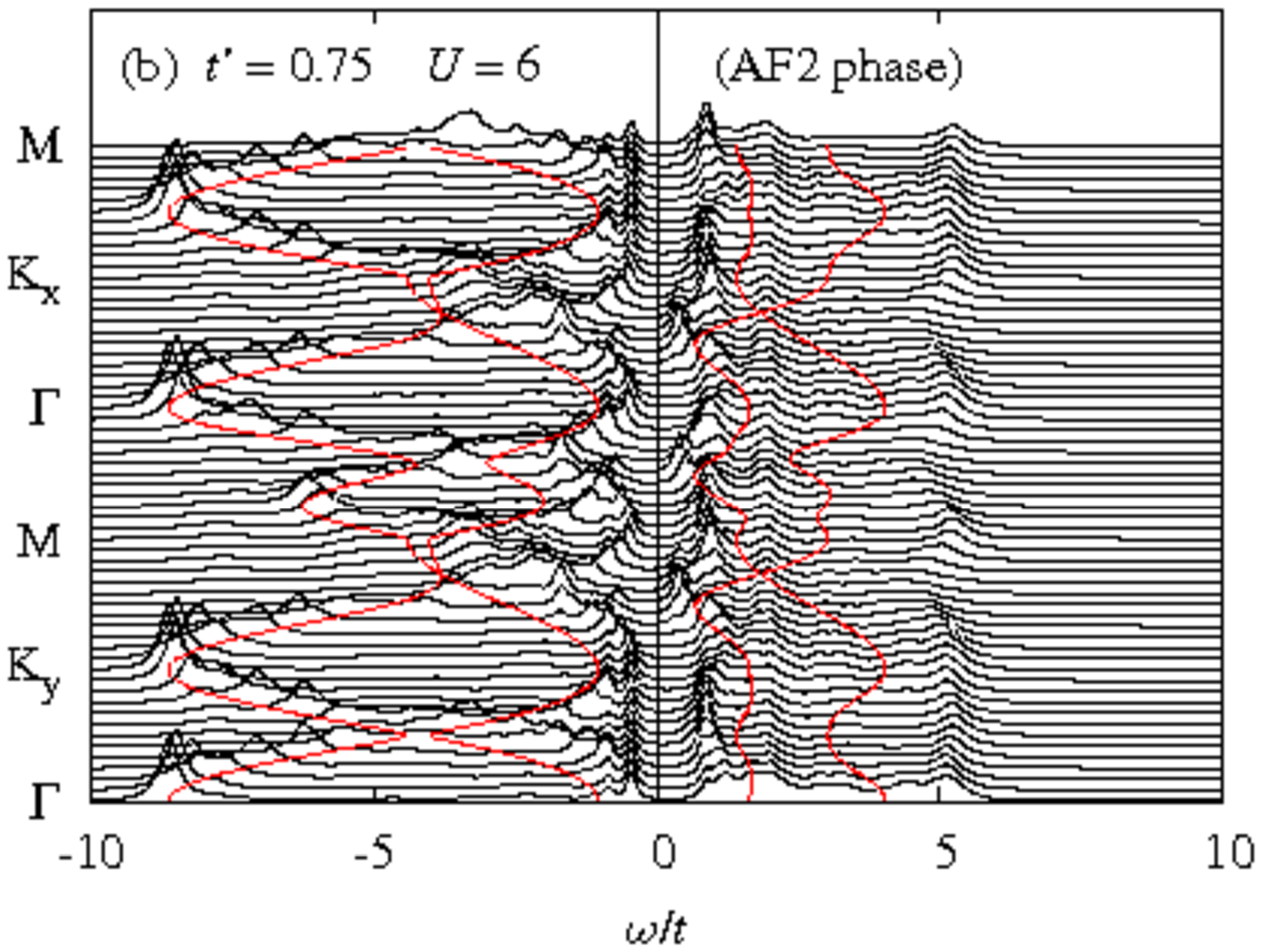}}
\subfigure{\includegraphics[width=0.47\textwidth, trim = 90 50 250 670, clip]
 {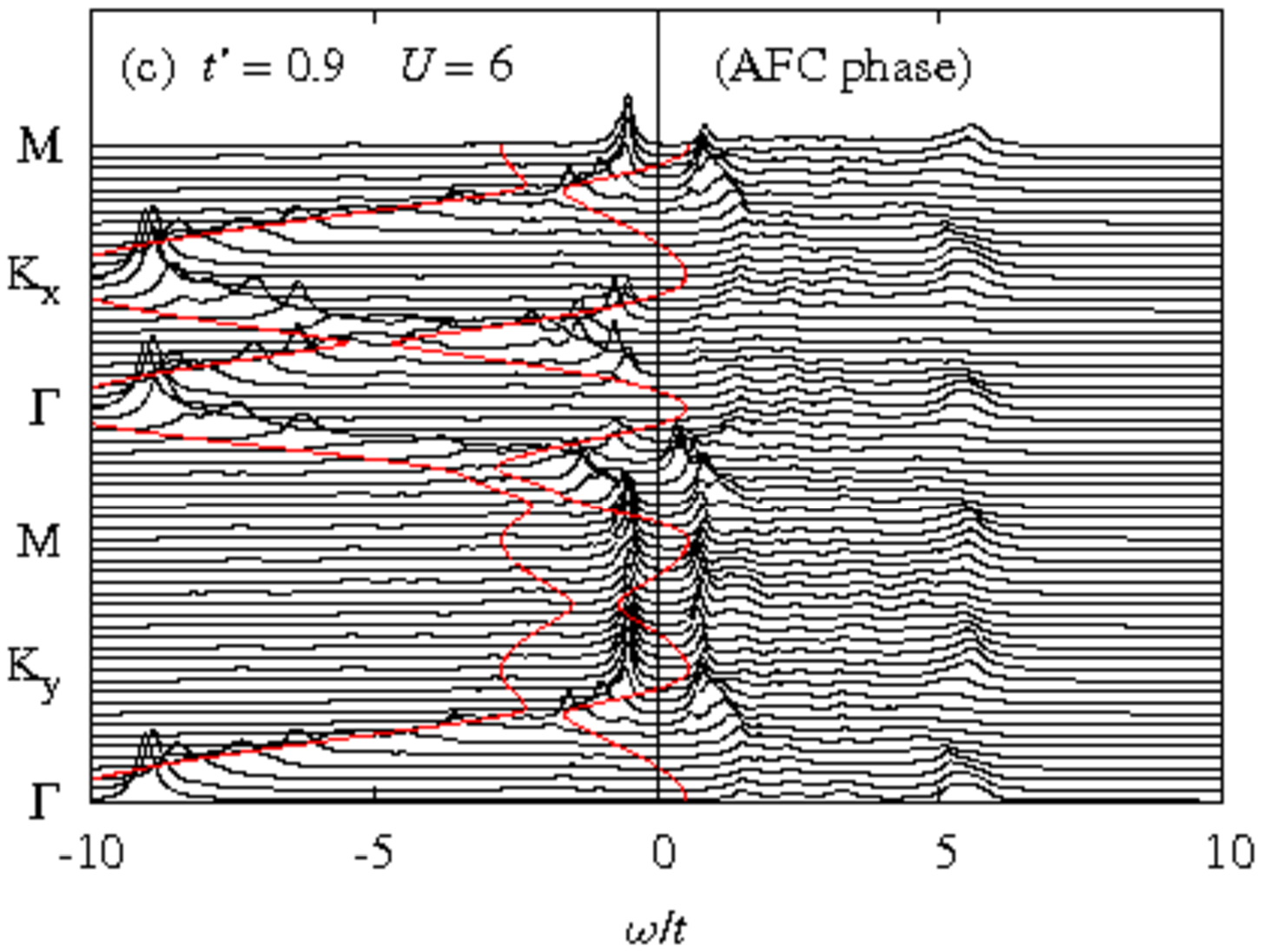}}
\caption{
(Color online) The spectral function of the 
(a) AF ($t' = 0.6$), (b) AF2 ($t'=0.75$), and (c) AFC ($t'=0.9$) solutions at $U=6$ 
calculated by VCA using the 3$\times$4 cluster. 
The Lorentzian broadening with $\eta = 0.1t$ is used in all cases. 
The mean-field dispersions are also plotted (full lines). 
\label{fig:sp-3x4-af-af2-afs}\\[-1.5em]
}
\end{figure}

Next we study in detail the spectral function by VCA using the 3$\times$4 cluster. 
This cluster yields results qualitatively similar to the mean-field approximation in the region $U \lesssim W$ 
and indicates the interesting possibility that the AF2 state is realized. 
It seems to us, though maybe subjective to some extent, that among the three clusters used here, 
the 3$\times$4 cluster simulates best the correlations in the bulk, and 
the fact that the stability of the AF2 phase is also obtained in the mean-field approximation 
suggests that it may not be due to some artifacts 
intrinsic to the 3$\times$4 cluster but is worth being further studied as an interesting possibility. 
\begin{figure}
\includegraphics[width=0.47\textwidth,trim = 0 0 0 0,clip]{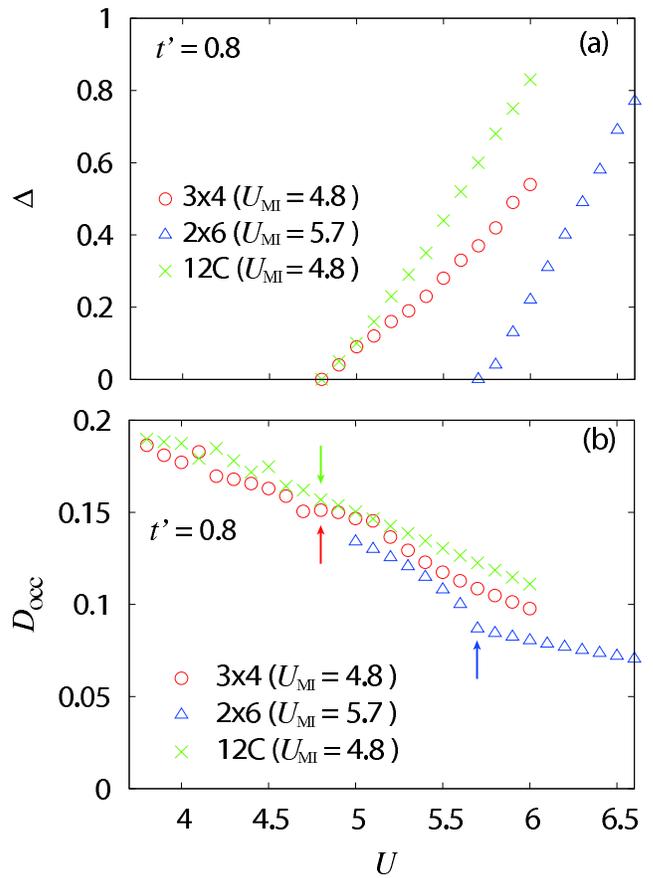}
\caption{
(Color online)
(a) The Mott gap $\Delta$ and (b) double occupancy $D_{\rm occ}$ as functions of $U$ at $t' = 0.8$ obtained by VCA 
on the 3$\times$4 (circles), 2$\times$6 (triangles), and 12C (crosses) clusters, assuming that no magnetic order 
is allowed. In (b) the Mott transition points are indicated by the arrows. 
\label{fig:gap-docc}\\[-1.5em]
}
\end{figure}

\begin{figure*}
\subfigure{\includegraphics[width=0.45\textwidth, trim = 70 80 250 680, clip]
 {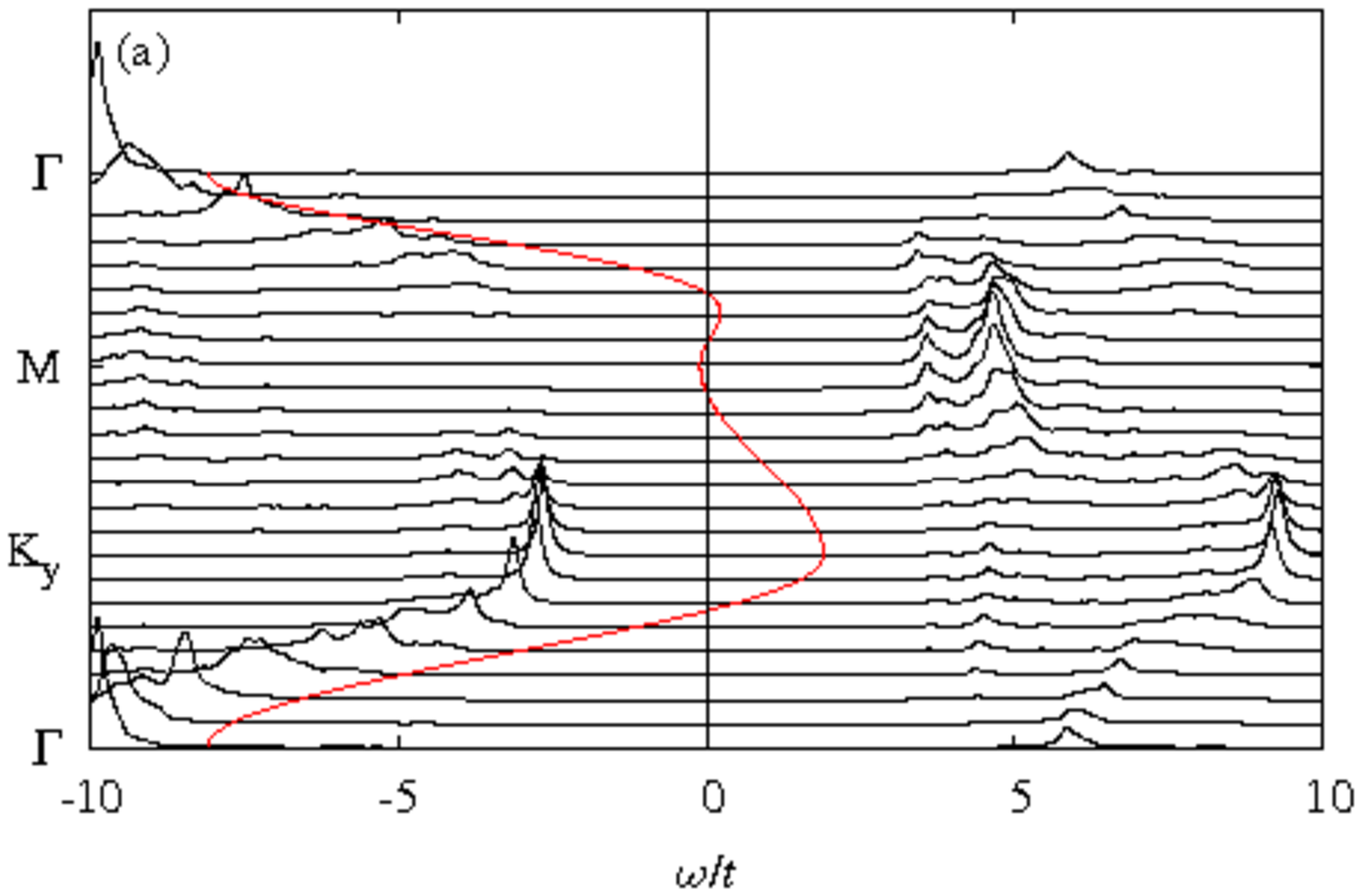}}
\subfigure{\includegraphics[width=0.45\textwidth, trim = 70 80 250 680, clip]
 {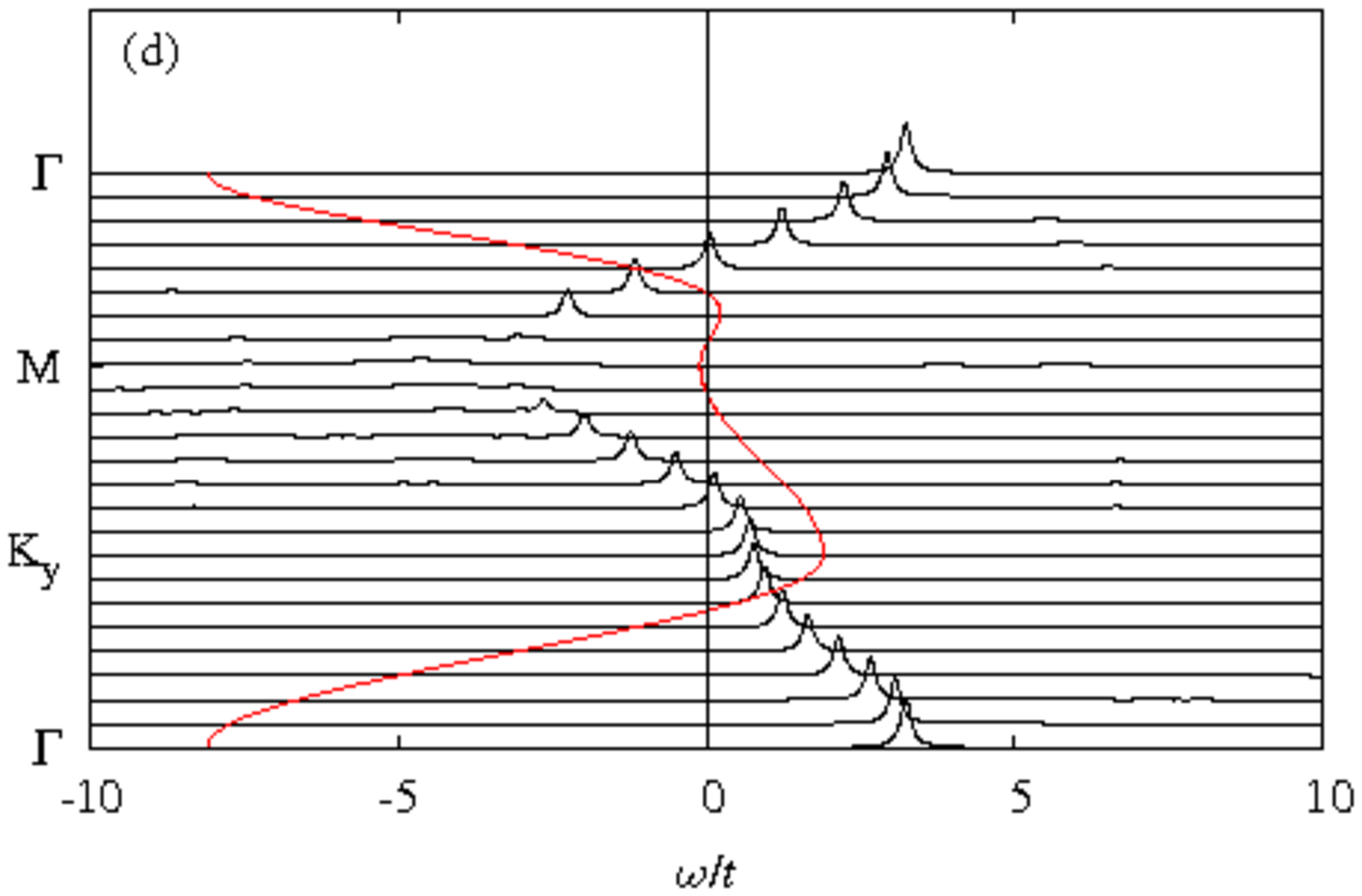}}
\subfigure{\includegraphics[width=0.45\textwidth, trim = 70 80 250 680, clip]
 {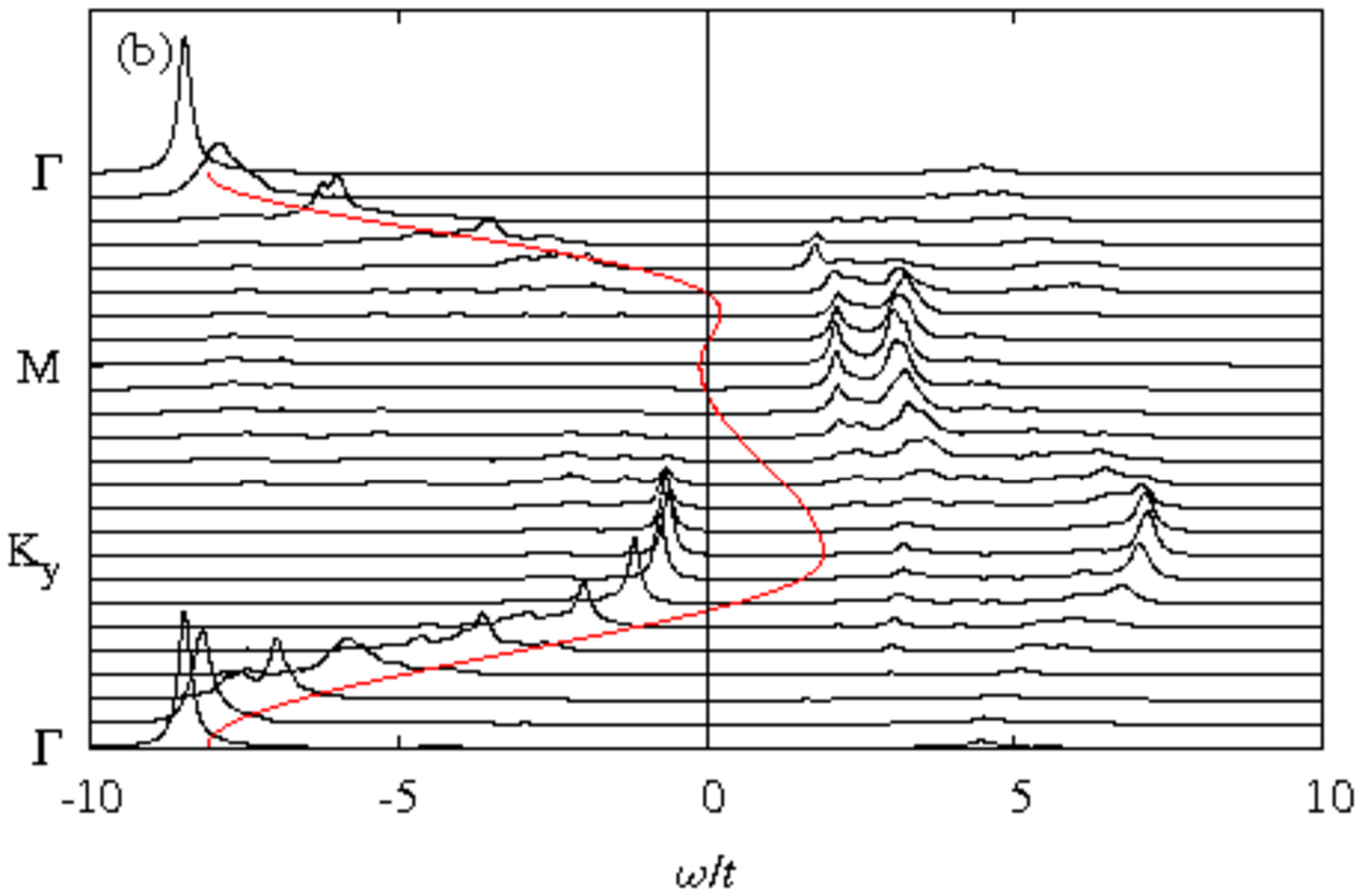}}
\subfigure{\includegraphics[width=0.45\textwidth, trim = 70 80 250 680, clip]
 {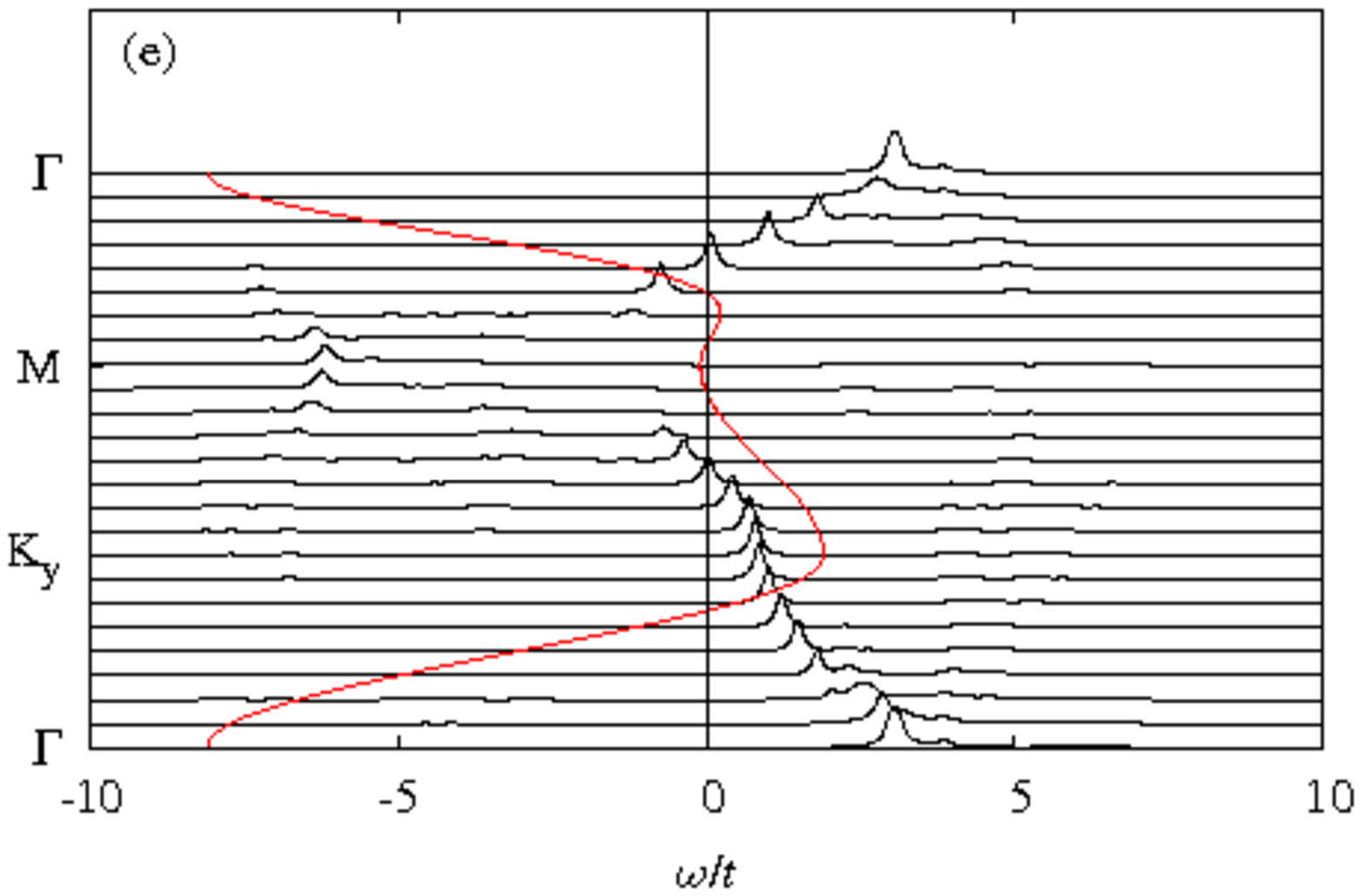}}
\subfigure{\includegraphics[width=0.45\textwidth, trim = 70 80 250 680, clip]
 {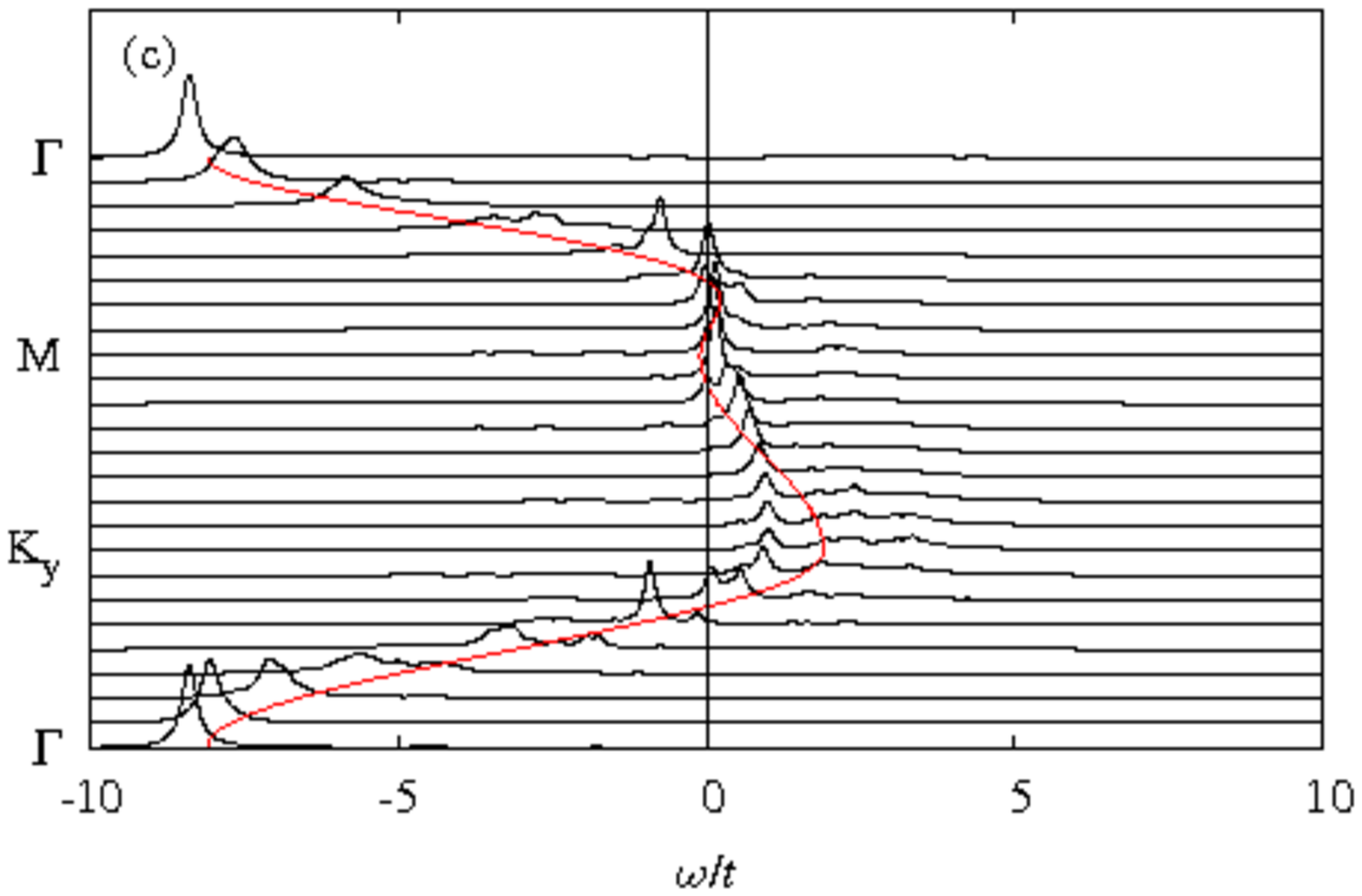}}
\subfigure{\includegraphics[width=0.45\textwidth, trim = 70 80 250 680, clip]
 {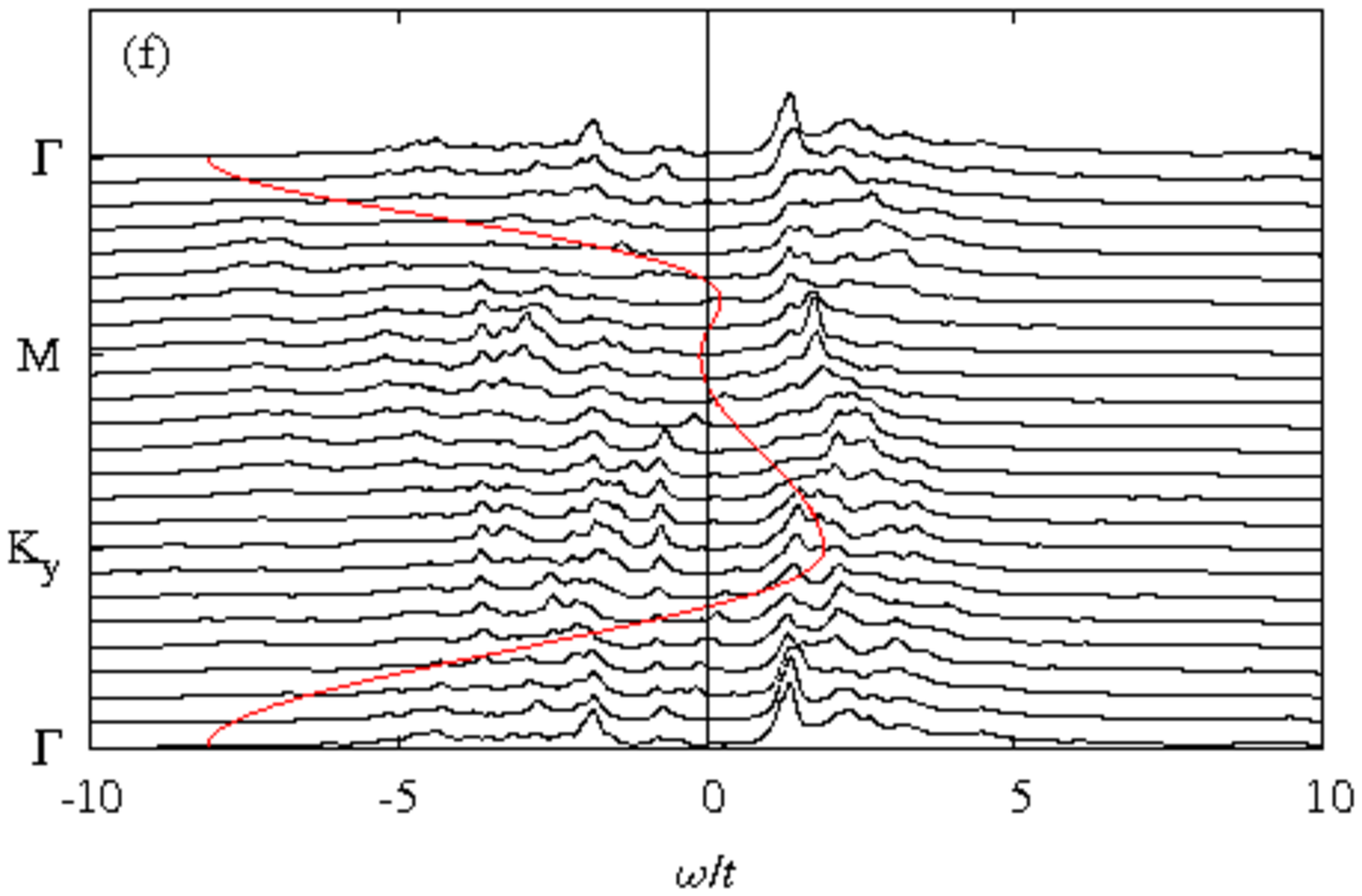}}
\caption{
(Color online) 
The spectral function $A(k,\omega)$ of the paramagnetic state for 
(a) $U=12$ (Mott insulator), (b) $U=8$ (Mott insulator), and (c) $U=4$ (metal) at $t'=0.75$
calculated by VCA using the 3$\times$4 cluster. 
The spectral function $A_{\Sigma}(k,\omega)$ of the self-energy is also calculated for the same 
$U$ and $t'$ ((d), (e), and (f)).
The Lorentzian broadening with $\eta = 0.1t$ is used in all cases. 
The full lines are the noninteracting band structure at the half-filling. 
\label{fig:3x4-para-sp-self-sp}\\[-1.0em]
}
\end{figure*}
In Fig.~\ref{fig:sp-3x4-af-af2-afs} we show the spectral function for the 
AF ($t' = 0.6$), AF2 ($t'=0.75$), and AFC ($t'=0.9$) solutions at $U=6$ 
calculated by VCA using the 3$\times$4 cluster, where 
the magnetic Brillouin zones are depicted in Fig.~\ref{fig:model} (b) 
and the $x$ and $y$ directions are indicated in Fig.~\ref{fig:spin}. 
The mean-field dispersions are also plotted (full lines). 
The order parameters computed by VCA are 
$M = 0.60$ for the AF, $M = 0.26$ for the AF2, and $M = 0.29$ for the AFC solutions, 
while the mean-field prediction is $M = 0.83$  for the AF and AF2, and AFC states. 
(The coincidence of the value $M$ for the three magnetic states is purely accidental.)
Though the quasiparticle dispersions in VCA are similar to the mean-field dispersions far from the Fermi level, 
they are largely modified near the Fermi level due to the electron correlations. 
The asymmetry of the spectral functions in the $x$ and $y$ directions is observed in the AF2 and AFC solutions. 
As is seen in Fig.~\ref{fig:model} (b), for example, 
$\Gamma \rightarrow {\rm K}_y$ part is identical to ${\rm M} \rightarrow  {\rm K}_y$ part for the AF state, 
it is identical to $(\pi,-\pi/2) \rightarrow (\pi,\pi/2)$ part for the AF2 state, and 
it is identical to ${\rm K}_x \rightarrow {\rm M}$ part for the AFC state. 

\subsection{Mott transition}

Next we study the Mott transition in detail. Fig.~\ref{fig:gap-docc} shows 
(a) the Mott gap $\Delta$ and (b) double occupancy $D_{occ}$ as functions of $U$ at $t' = 0.8$ obtained by VCA 
on the 3$\times$4 (circles), 2$\times$6 (triangles), and 12C (crosses) clusters, assuming that no magnetic order 
is allowed (i.e. $h_{\rm M} = 0$). 
The Mott gap closes continuously at $U=U_{\rm MI}$ and below $U_{\rm MI}$ 
there are no energetically disfavored Mott insulator solutions, so the Mott transition is of the second order. 
In general, Mott transitions are predicted to be first order in the variational cluster approach 
with bath degrees of freedom where hybridization between the bath sites and cluster sites is treated 
as a variational parameter.\cite{Potthoff2,Potthoff3} 
In these analyses, the coexisting metal and insulator solutions, leading to the first order transition, 
differ by the value of these hybridization parameters. 
Our analysis does not have bath degrees of freedom and technically this will be the origin of the difference. 
It remains to be clarified which is the correct picture. 

Next we study in detail the spectral densities and related quantities.
The spectral density is defined by 
\begin{eqnarray}
A(k\sigma,\omega) = - \frac{1}{\pi} {\rm Im} G(k\sigma, \omega  + i \eta)
\end{eqnarray}
where the Green function $G(k\sigma,z)$ is expressed as 
\begin{eqnarray}
G(k\sigma, z ) = \frac{1}{z -  \varepsilon_{k} - \Sigma(k\sigma,z) }
\label{eqn:g-sigma}
\end{eqnarray}
in terms of the self-energy $\Sigma(k\sigma,z)$, whose spectral representation is\cite{l1} 
\begin{eqnarray}
\Sigma(k\sigma,z) = g_{k\sigma} + \int_{- \infty}^{\infty} \frac{\sigma_{k\sigma}(\xi)}{z-\xi},
\,\,\,\,\,\sigma_{k\sigma}(\xi) \geq 0.
\label{eqn:sigma-spectral-rep}
\end{eqnarray}
Useful information on $g_{k\sigma}$ and $\sigma_{k\sigma}(\xi)$ 
can be obtained\cite{eder1} by comparing the $|z| \rightarrow \infty$ expansions of the 
Eq. (\ref{eqn:g-sigma}) with (\ref{eqn:sigma-spectral-rep}) and that of the expression
\begin{eqnarray}
G(k\sigma, z ) &=&
\langle \Omega  | \{ 
c_{k\sigma }  \frac{1}{z - H  + E_0}   c_{k\sigma }^\dag  \nonumber \\ 
& &+ c_{k\sigma }^\dag  \frac{1}{z + H  - E_0} c_{k\sigma }
\}  | \Omega \rangle,
\end{eqnarray}
where $| \Omega \rangle $ and $E_0$ are the ground state and ground-state energy of $H$, 
and for the Hamiltonian (\ref{eqn:hm}), 
\begin{eqnarray}
g_{k\sigma} = U\langle n_{-\sigma } \rangle,\,\,
\int_{- \infty}^{\infty} \sigma_{k\sigma}(\xi) = U \langle n_{-\sigma } \rangle (1 - \langle n_{-\sigma } \rangle),
\nonumber
\end{eqnarray}
where $\langle n_{\sigma } \rangle $ is the average number per site of electrons with spin $\sigma$ 
in the ground state $| \Omega \rangle $.

Here we consider the paramagnetic state at half-filling and set $\langle n_{\pm \sigma } \rangle = 1/2$. 
When the spectral weight $\sigma_{k\sigma}(\xi)$ becomes dominated by a single pole of the dispersion $\xi_k$, 
the Green function is given by 
\begin{eqnarray}
G(k\sigma, z ) = \frac{1}{z - \tilde{\varepsilon}_{k} - \frac{ U^2/4}{z-\xi_k}  },
\label{eqn:g-sigma-approx}
\end{eqnarray}
where $\tilde{\varepsilon}_{k} = \varepsilon_{k} + g_{k\sigma}$. 
When $U$ is large, $g_{k\sigma} = U/2$ is cancelled by the chemical potential $\mu \simeq U/2$ 
in $\varepsilon_{k}$ at half-filling, so $\tilde{\varepsilon}_{k}$ becomes more or less the same as the dispersion of the 
metallic state. Also, the atomic limit $t=t'=0$ yields the self-energy of the form in Eq. (\ref{eqn:g-sigma-approx}) 
with $\xi_{k} = 0$, which suggests that $\xi_{k}$ will be almost independent of $U$ for large $U$. 
Therefore, when $U$ is large so that $ U \gg |\tilde{\varepsilon}_{k} |, |\xi_k | $, 
the poles of the Green function (\ref{eqn:g-sigma-approx}) are given by 
\begin{eqnarray}
\omega_{\pm} = \frac{1}{2}( \tilde{\varepsilon}_{k} + \xi_k ) \pm \frac{U}{2},
\label{eqn:poles}
\end{eqnarray}
thus the original band $\tilde{\varepsilon}_{k}$ splits into the upper and lower Hubbard bands (of almost equal weights) 
and a gap of width $U$ opens. 
As is observed here, the spectral density of the self-energy $\Sigma(k\sigma,z)$, 
\begin{eqnarray}
A_{\Sigma}(k\sigma,\omega) = - \frac{1}{\pi} {\rm Im} \Sigma(k\sigma, \omega  + i \eta)
\end{eqnarray}
is a key to understand the Mott transition, which we study in detail together with the spectral density $A(k\sigma,\omega)$.

In Fig.~\ref{fig:3x4-para-sp-self-sp} we show 
the spectral function $A(k,\omega)$ of the paramagnetic state for 
(a) $U=12$ (Mott insulator), (b) $U=8$ (Mott insulator), and (c) $U=4$ (metal) at $t'=0.75$
calculated by VCA using the 3$\times$4 cluster. 
The spectral function $A_{\Sigma}(k,\omega)$ of the self-energy is also calculated for the same $U$ and $t'$ ((d), (e), and (f)).
The noninteracting band structure is plotted with full lines. 
Observing e.g. the region around ${\rm K}_y$ the band shrinks toward the Fermi level 
compared to the noninteracting case in the metallic phase (c), 
and it splits into upper and lower Hubbard bands in the Mott insulator state (b) and the gap opens. 
Below the Mott transition point ($U=4$) the $A_{\Sigma}(k,\omega)$ seems to be almost featureless, but it becomes 
dominated by the single pole of well-defined definite dispersion above the Mott transition point ($U=8$) and, 
comparing (d) and (e) the dispersion is almost independent of $U$, as it is expected. 

\begin{figure}
\includegraphics[width=0.47\textwidth,trim = 0 0 0 0,clip]{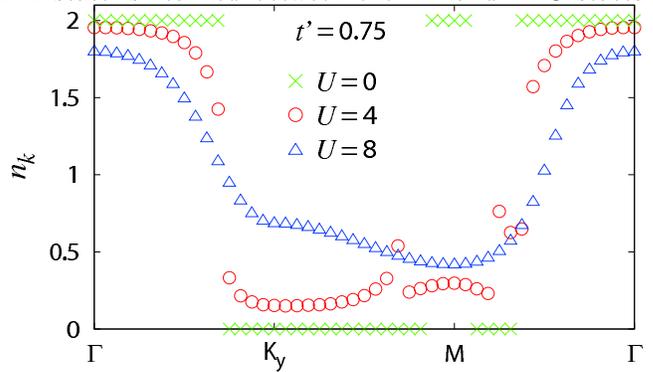}
\caption{
(Color online) 
The momentum distribution function $n_k$ for $U = 0$ (crosses), $U=4$ (circles), and $U=8$ (triangles) at $t'=0.75$
calculated by VCA using the 3$\times$4 cluster. 
\label{fig:nk}
}
\end{figure}

Supplementing these analyses, we show in Fig.~\ref{fig:nk} the momentum distribution function $n_k$ 
for $U=0$ (crosses) , $U=4$ (circles), and $U=8$ (triangles) at $t'=0.75$ calculated by VCA using the 3$\times$4 cluster. 
The sharp gap of $n_k$ at the Fermi surface in the metal disappears in the Mott insulator phase.

\section{Summary and conclusions}
\label{sec:summary}

We have studied the magnetic properties and Mott transition in the Hubbard model on the square lattice 
by VCA and analyzed the phase diagram at half-filling and zero temperature. 
12-site clusters of three different shapes (3$\times$4, 2$\times$6, and 12C cluster) are used to see the 
finite-size effects of the analysis. 

In the region $ U \gtrsim  W (=8) $, the predictions of the three clusters agree well, and 
the AF state is realized for $t' \lesssim 0.65$, while $0.8 \lesssim  t'$ the AFC state is stable, 
and the PM state is realized between the AF and AFC states. 
There is no region where the incommensurate AF2 phase, 
suggested in the previous study,\cite{mizusaki} is stable for $ U \gtrsim  W $. 

In the strong coupling limit, our results quantitatively agree well with the analysis of the 
Heisenberg model, which supports the validity of our analysis in the region $ U \gtrsim  W$. 

In the region $ U \lesssim W $, the VCA on the 3$\times$4 cluster 
yields the results qualitatively very similar to the mean-field analysis, and shows 
that the AF2 state is realized for $0.7 \lesssim t'/t \lesssim 0.85$ and $4.5 \lesssim U \lesssim 6.5$, though 
the cluster-shape dependence of the results are not negligible. 
So incommensurate states may be realized, but in the limited area around this region.

The Mott transition takes place slightly below the band width $W$ and is of the second order. 
The spectral density of the self-energy is featureless below the Mott transition point (metallic phase), 
but in the Mott insulator phase, it becomes dominated by single pole of a definite dispersion, which yields the Mott gap. 
A detailed study on the nature of the PM state between the AF and AFC states, such as the dimer phase, is 
an interesting subject, but it is unfortunately beyond the scope of VCA. 

\section{Acknowledgement}

One of us (A.Y.) would like to thank K.~Kurasawa and H.~Nakada for useful discussions on numerical analysis. 
Parts of numerical calculations were done using the computer facilities of the IMIT at Chiba University, 
the Yukawa Institute, and the Research Center for Computational Science at Okazaki, Japan. 
This work is supported in part by Kakenhi Grant No. 22540363 of Japan.

\end{document}